\documentclass[12pt,preprint]{emulateapj}
\newcommand{\be}{\begin{equation}}
\newcommand{\ee}{\end{equation}}
\newcommand{\ba}{\begin{eqnarray}}
\newcommand{\ea}{\end{eqnarray}}

\begin{document}

\title{Probing the Pre-Reionization Epoch with Molecular Hydrogen Intensity Mapping}

\author{Yan Gong$^1$, Asantha Cooray$^1$, Mario G. Santos$^2$}

\affil{$^1$Department of Physics \& Astronomy, University of California, Irvine, CA 92697}
\affil{$^2$CENTRA, Instituto Superior T\'ecnico, Technical University of Lisbon, Lisboa 1049-001, Portugal}

\begin{abstract}
Molecular hydrogen is now understood to be the main coolant of the primordial gas clouds 
leading to the formation of the very first stars and galaxies. 
The line emissions associated with molecular hydrogen
should then be a good tracer of the matter distribution at the onset of reionization of the universe. Here we 
propose intensity mapping of $\rm H_2$ line emission in rest-frame mid-infrared wavelengths 
to map out the spatial distribution of gas at redshifts $z > 10$.
We calculate the expected mean intensity and clustering power spectrum for
several $\rm H_2$ lines. We find that the 0-0S(3) rotational line at a rest wavelength of $9.66$ $\rm \mu m$ is 
the brightest line over the redshift range of 10 to 30 with an intensity of about 5 to 10 Jy/sr at $z\sim15$. 
To reduce astrophysical and instrumental systematics, we propose the cross-correlation between 
multiple lines of the H$_2$ rotational and vibrational 
line emission spectrum. Our estimates of the intensity  can be used as a guidance in planning instruments 
for future mid-IR spectroscopy missions such as SPICA. 
\end{abstract}


\maketitle

\section{Introduction}
Existing cosmological observations show that the reionization history of the universe at $z > 6$
is likely both complex and inhomogeneous (e.g. Haiman 2003; Choudhury \& Ferrara 2006; Zaroubi 2012). 
While the polarization signal in the cosmic microwave background (CMB) anisotropy power
spectrum constrains the total optical depth to electron scattering and the existing WMAP measurements suggest
reionization happened around $z_{\rm ri}=11$ (Komatsu et al. 2011), it is more likely that the reionization period 
was extended over a broad range of redshifts from 20 to 6. Moving beyond CMB,
observations of the 21-cm spin-flip line of neutral hydrogen are now pursued
to study the spatial distribution of the matter content during the epoch of reionization (e.g., Madau et
al. 1997; Loeb \& Zaldarriaga 2004; Gnedin \& Shaver 2004). Unlike CMB, 21 cm data are
useful as they provide a tomographic view of
the reionization (Furlanetto et al. 2004; Santos et al. 2005). 
The anisotropy power spectrum of the 21 cm line emission is also a 
useful cosmological probe (Santos \& Cooray 2006; McQuinn et al. 2006; Bowman et al. 2007; Mao et al. 2008).

While the 21 cm signal is  primarily tracing the neutral hydrogen content in
the intergalactic medium during reionization, line emission associated with atomic and molecular
lines are of interest to study the physical properties within dark matter halos, such as gas cooling, star-formation,
and the spatial distribution of first stars and galaxies. Motivated by various experimental possibilities we have 
studied the reionization signal associated with the CO (Gong et al. 2011), CII (Gong et al. 2012), and
Lyman-$\alpha$ (Silva et al. 2012) lines. As the signal is sensitive to the metal abundance these atomic and
molecular probes are more sensitive to the late stages of reionization, perhaps 
well into the epoch when the universe is close to full reionization and
has a low 21 cm signal (Basu et al. 2004; Righi et al. 2008; Visbal \& Loeb 2010; Carilli 2011; Lidz et al. 2011). 

Although the end of reionization era can be effectively probed with HI, CO, CII, and Lyman-$\alpha$,
it would be also useful  to have an additional probe of the onset
of reionization at $z > 10$. Here we consider molecular hydrogen and study the
signal associated with rotational and vibrational lines in the mid-IR wavelengths. 
Molecular hydrogen has been invoked as a significant coolant
of primordial gas leading to the formation of first stars and galaxies (e.g. Haiman 1999; Bromm \& Larson 2004; Glover 2005; Glover 2012). While molecular hydrogen is easily destroyed in later stages
of reionization, its presence in the earliest epochs of the cosmological history can be 
probed with line emission experiments.

This paper is organized as follows: in the next section, we outline the calculation related to the 
cooling rate of $\rm H_2$ rotational and vibrational lines. In Section 3  we
present results on the H$_2$ luminosity as a function of the halo mass, while in Section 4, we discuss the
mean H$_2$ intensity and clustering auto and cross power spectra. The cross power spectra between various lines
are proposed as a way to eliminate the low-redshift contamination and increase the overall 
signal-to-noise ratio for detection. We discuss potential detectability in Section~5.
We summarize our results and conclude in Section~6.
We assume the flat $\Lambda$CDM with $\Omega_{\rm M}=0.27$, $\Omega_{\rm b}=0.046$, $\sigma_8=0.81$, 
$n_{\rm s}=0.96$ and $h=0.71$ for the calculation throughout the paper (Komatsu et al. 2011).

\section{$\rm H_2$ cooling coefficients}

The radiation emitted by $\rm H_2$ will be generated by the heating/cooling of the gas as the collapsing process evolves.
Therefore, in order to calculate the $\rm H_2$ luminosity, 
we first evaluate the cooling rate of the H$_2$ rotational and vibrational lines
for optical thin and optical thick media, respectively. 
For hydrogen density $n_{\rm H}<10^9 \rm cm^{-3}$, the optical depth is thin for $\rm H_2$ 
emission lines. Following Hollenbach \& McKee (1979), the cooling coefficient for the 
rotational and vibrational lines can then be expressed as
\be\label{eq:H2_cooling}
\Lambda_{\rm H_2}^{\rm r,v}({\rm H,H_2}) = \frac{\Lambda_{\rm LTE}^{\rm r, v}({\rm H,H_2})}{1+n_{\rm cr}^{\rm r,v}({\rm H,H_2})/n_{\rm H,H_2}},
\ee
where $\Lambda_{\rm LTE}^{\rm r, v}({\rm H,H_2})$ is the cooling coefficients of rotation or vibration at the 
local thermodynamic equilibrium (LTE) for the collisions with hydrogen atoms H or $\rm H_2$.  Here
$n_{\rm cr}^{\rm r,v}$ is the critical density of H or $\rm H_2$ to reach the LTE, and 
$n_{\rm H,H_2}$ is the local number density of H or $\rm H_2$.

In the LTE we have $A_{\rm ul}=C_{\rm ul}n$ where $A_{\rm ul}$ is the Einstein coefficient, $C_{\rm ul}$ 
is the collisional de-excitation rate from upper to lower level, and $n$ is the particle number density 
(Hollenbach \& McKee 1979). Then the rotational and vibrational LTE cooling coefficient can be written as
\be \label{eq:L_LTE_rv}
\Lambda_{\rm LTE}^{\rm r,v}({\rm H,H_2}) = \frac{1}{n_{\rm H,H_2}}A_{J}\frac{g_J}{g_{J'}}e^{-\frac{\Delta E}{kT}}\Delta E,
\ee
where $g_J=2J+1$ is the statistical weight, $J$ denotes the total angular momentum quantum number 
of the rotational energy level, $A_{J}$ is the Einstein coefficient for $J \to J'$ transition at the
same vibrational energy level or between two vibrational energy levels,
and $\Delta E$ is the energy difference between $E_{J}$ and $E_{J'}$. The wavenumber for each energy level
and the calculation for $E_J$ and wavelength are given in the Appendix. 
The values of $A_J$ are taken from Turner et al. (1977). 
Note that we just consider two vibrational level transition v=0 and 1 in the following calculation.
The number of H$_2$ at the higher vibrational levels are much smaller than that at v=0 and 1 in our case, 
so the strength of these lines is much weaker than those from v=0 and 1.
The $n_{\rm cr}/n$ term in Eq.(\ref{eq:H2_cooling}) can be approximated by (Hollenbach \& McKee 1979)
\be
\frac{n_{\rm cr}^{\rm r,v}({\rm H,H_2})}{n_{\rm H,H_2}} = \frac{\Lambda_{\rm LTE}^{\rm r,v}({\rm H,H_2})}{\Lambda_{\rm n\to 0}^{\rm r,v}({\rm H,H_2})},
\ee
where $\Lambda_{\rm n\to 0}^{\rm r,v}$ is the low-density limit of the cooling coefficient, which can be
obtained by replacing the Einstein coefficient in Eq.(\ref{eq:L_LTE_rv})
by $C_J^{\rm H,H_2}n_{\rm H,H_2}$, i.e.
\be \label{eq:L_n0}
\Lambda_{\rm n\to 0}^{\rm r,v}({\rm H,H_2}) = C_J^{\rm H,H_2}\frac{g_J}{g_{J'}}e^{-\frac{\Delta E}{kT}}\Delta E.
\ee
Here $C_J^{\rm H,H_2}$ is the collisional de-excitation coefficients with H or H$_2$ for $J \to J'$ transition 
in the same vibrational level or between v=$i$ and v=$j$, which are estimated by the fitting 
formula given in Hollenbach \& McKee (1979) and Hollenbach \& McKee (1989) (see Appendix). 
Using Eq.(\ref{eq:H2_cooling})-(\ref{eq:L_n0}) we can estimate the cooling coefficient for a given 
$\rm H_2$ line. Note that in Hollenbach \& McKee (1979) the fitting formulae of {\it total} cooling 
coefficients for v=0 rotational  and v=0, 1 and 2 vibrational lines are given.  We denote those by
$\Lambda_{\rm H_2}^{\rm tot, rv}$, $\Lambda_{\rm LTE}^{\rm tot,rv}$ and $\Lambda_{n \to 0}^{\rm tot,rv}$ 
in the cases of the total local, LTE and low-density limit cooling coeffecients, respectively. 
Then $n_{\rm cr}/n$ can be expressed by $\Lambda_{\rm LTE}^{\rm tot,rv}/\Lambda_{n \to 0}^{\rm tot,rv}$, 
where $n_{\rm cr}$ denotes the critical density when all energy level transitions are in the LTE.

For the hydrogen density $n_{\rm H}>10^9 \rm cm^{-3}$, the optical depth is thick for the $\rm H_2$ emission, 
and we have to consider the absorption effect for the $\rm H_2$ cooling. Following Yoshida et al. (2006), 
we make use of the cooling efficiency, which is defined by $f_{\rm ce}=\Lambda_{\rm thick}/\Lambda_{\rm thin}$, 
to evaluate the cooling coefficient $\Lambda_{\rm thick}$ for the optical thick case. This reduction factor is 
derived from their simulations, and available for $n_{\rm H}\lesssim 10^{14} \rm cm^{-3}$, which is well within the
density ranges of our calculation. The $f_{\rm ce}$ is about 0.02 when $n_{\rm H}\sim 10^{14} \rm cm^{-3}$ and
increases to about 0.1 when $n_{\rm H}\sim 10^{12} \rm cm^{-3}$. At densities below $10^{10} \rm cm^{-3}$ we 
have $f_{\rm ce} = 1$ (Yoshida et al. 2006).

\begin{figure}[htb]
\includegraphics[scale = 0.4]{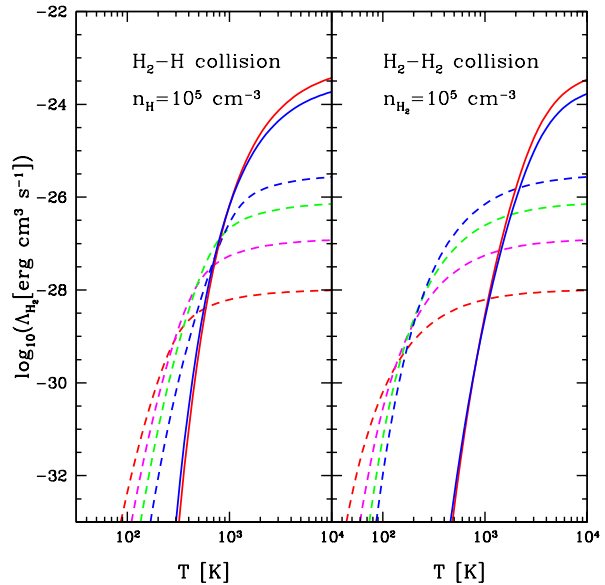}
\caption{\label{fig:H2_cooling} The optical-thin $\rm H_2$ cooling coefficient vs. 
temperature for $\rm H_2$-H and $\rm H_2$-$\rm H_2$ collisions at $z=15$. 
The cooling coefficients of the rotational lines, 0-0S(0) (red), 0-0S(1) (magenta),
0-0S(2) (green) and 0-0S(3) (blue), are shown in dashed curves, and the vibrational
lines, 1-0S(1) (red) and 1-0Q(1) (blue), are in solid curves.
}
\end{figure}

In Fig.~\ref{fig:H2_cooling}, as an example, we show the cooling coefficients for $\rm H_2$-H and 
$\rm H_2$-$\rm H_2$  collisions as a function of temperature $T$. We assume the number density of hydrogen atom H and
molecular hydrogen H$_2$ to be $10^{5}$ $\rm cm^{-3}$ here. The dashed curves show the cooling 
coefficients of the rotational lines, which are in red (0-0S(0)), magenta (0-0S(1)), green (0-0S(2)) and 
blue (0-0S(3)). The solid curves are for two vibrational lines 1-0S(1) in red and 1-0Q(1) in blue, respectively.
As can be seen, the rotational cooling dominates at the low temperature 
($T\lesssim 10^3$ K) and the vibrational cooling dominates at high temperature ($T\gtrsim 10^3$ K).
Also, we find the $\Lambda_{\rm H_2}^{\rm r}$ for $\rm H_2$-$\rm H_2$ collision is generally greater than 
that for $\rm H_2$-H collision at $T\lesssim 10^3$ K, while the $\Lambda_{\rm H_2}^{\rm v}$ for 
$\rm H_2$-$\rm H_2$ collision is less than that for $\rm H_2$-H collision in this temperature range.
This indicates that, at low temperature, the total $\Lambda_{\rm H_2}^{\rm r}$ is mainly from 
$\rm H_2$-$\rm H_2$ collisions, and  the total $\Lambda_{\rm H_2}^{\rm v}$ is from $\rm H_2$-H 
collisions. At higher temperature with $T\gtrsim 10^3$ K, the cooling rates
$\Lambda_{\rm H_2}^{\rm r}$ and $\Lambda_{\rm H_2}^{\rm v}$ 
for both $\rm H_2$-$\rm H_2$ and $\rm H_2$-H collisions are similar.

\section{$\rm H_2$ luminosity}

\begin{figure*}[htbp]
\centerline{
\includegraphics[scale = 0.4]{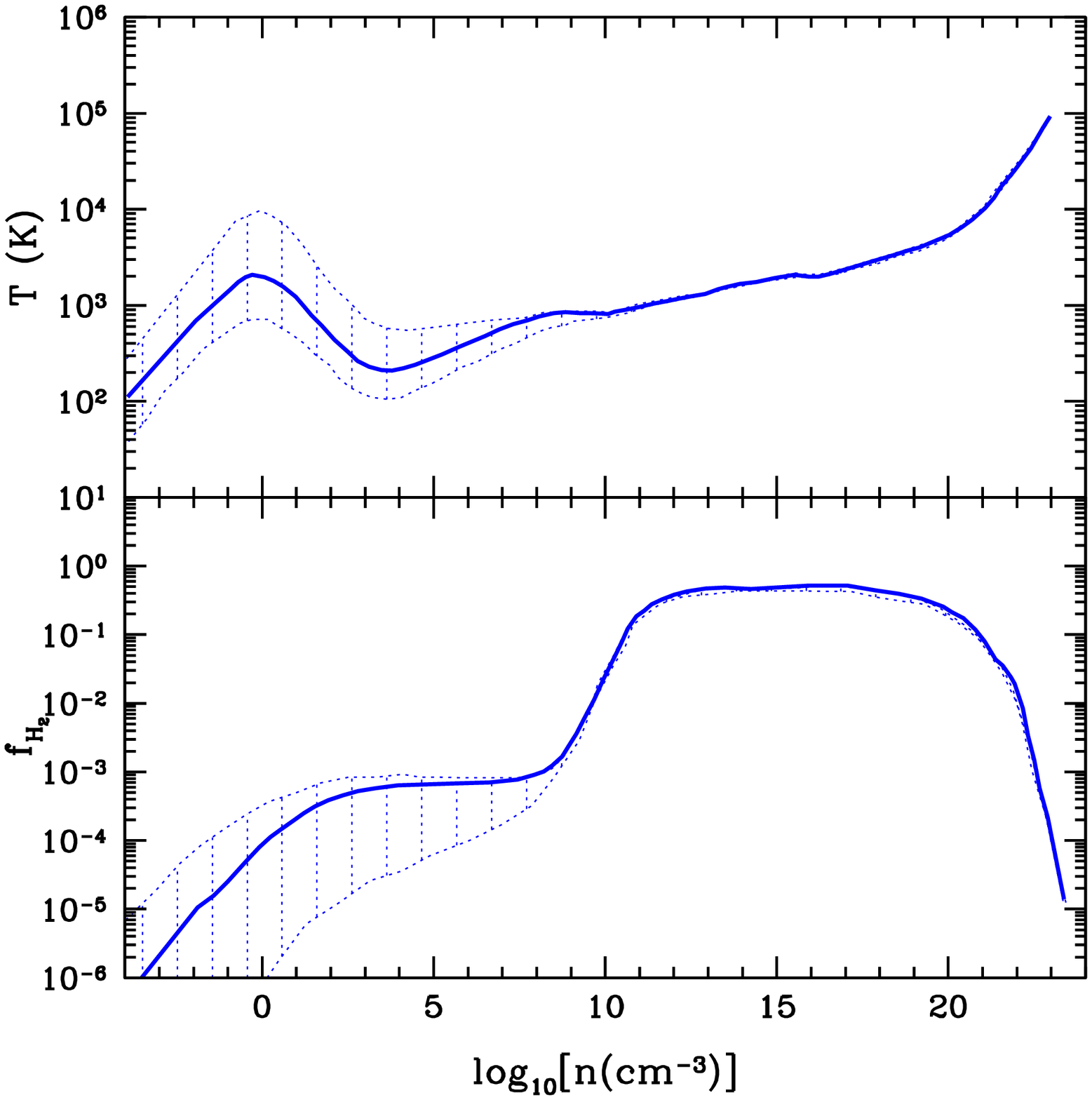}
\includegraphics[scale = 0.4]{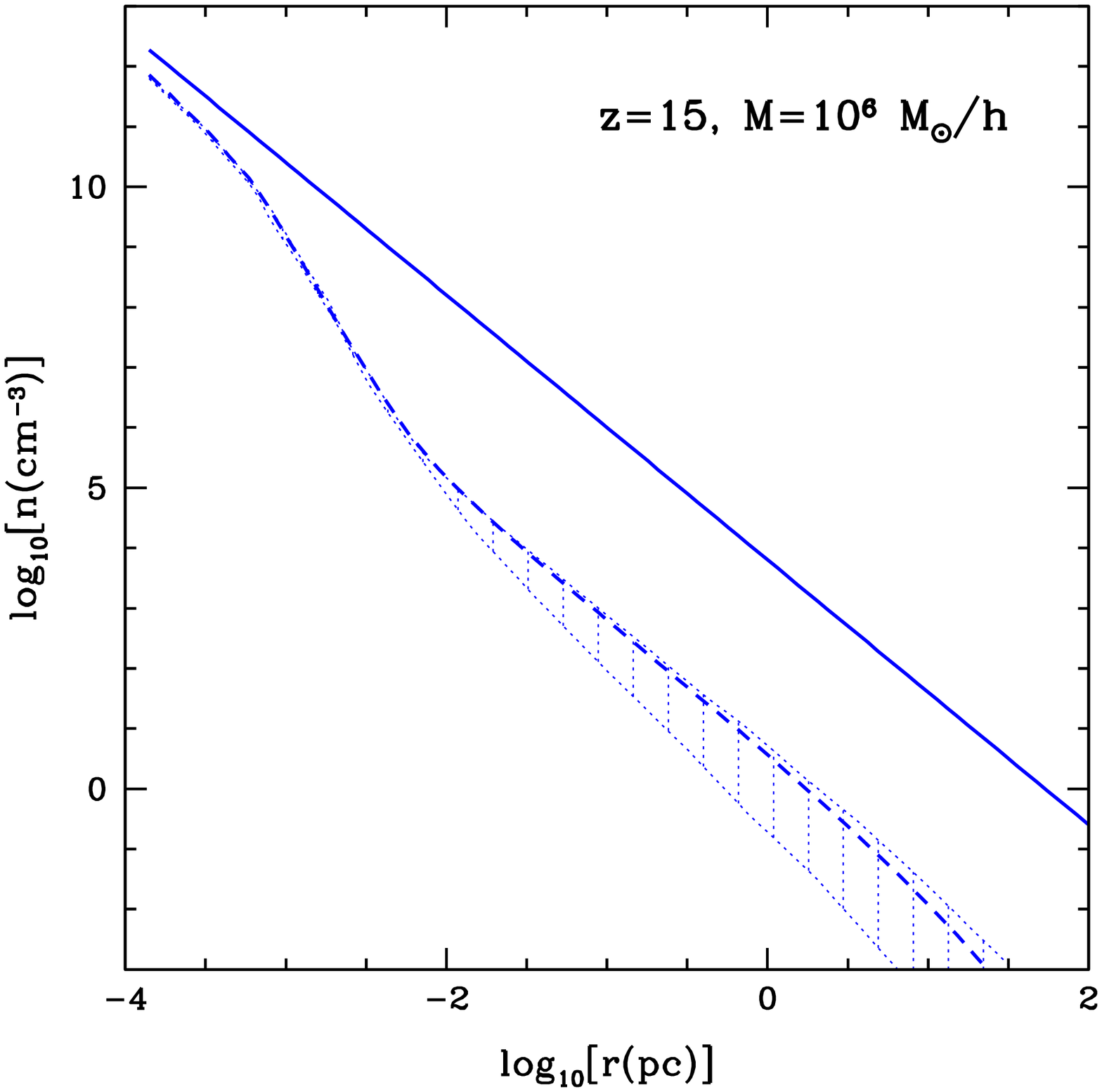} 
}
\caption{\label{fig:H2_gas}  
{\it Left}: The gas temperature $T$ and $\rm H_2$
fraction $f_{\rm H_2}$ as functions of the gas density $n$, which are derived
from the simulation results in Omukai (2001) and Yoshida et al. (2006). The
uncertainties of the gas temperature and H$_2$ fraction are shown in blue regions.
{\it Right:} The density profile of the gas (blue solid line) and molecular 
hydrogen (blue dashed line) for the halo with $M=10^6\ \rm M_{\sun}/h$ at $z=15$.
The blue region shows the uncertainty of the H$_2$ density profile estimated by
the uncertainty of $f_{\rm H_2}$ in the left panel.
}
\end{figure*}

We now explore the H$_2$ luminosity as a function of halo mass for rotational and vibrational lines.
As discussed in the last section, the $\rm H_2$ cooling coefficient $\Lambda_{\rm H_2}$ is dependent on the 
local gas temperature and density of hydrogen and molecular hydrogen. To evaluate 
$\rm H_2$ luminosity vs. halo mass relation for molecular hydrogen cooling within
primordial dark matter halos, we first need to know the radial profile of the gas temperature 
and density within dark matter halos.

Following the results from numerical simulations involving the 
formation of primordial molecular clouds (e.g. Omukai \& Nishi 1998; 
Abel et al. 2000; Omukai 2001; Yoshida et al. 2006; McGreer \& Bryan 2008), we assume the gas density 
profile as
\be \label{eq:p_gas}
\rho(r) = \rho_0\left(\frac{r}{r_0}\right)^{-2.2},
\ee
where we set $r_0=1$ pc and $\rho_0$ is the normalization factor which is obtained by
\be
M_{\rm gas}=4\pi \int_0^{r_{\rm vir}} r^2 \rho(r) dr \, .
\ee
Here $M_{\rm gas}=(\Omega_b/\Omega_M) M$ is the gas mass in the virial radius of the halo 
with dark matter mass $M$, and the $r_{\rm vir}$ is the virial radius which is given by
\be
r_{\rm vir} = \left[ \frac{M}{(4/3)\pi \rho_{\rm vir}} \right]^{1/3} \, .
\ee 
Here $\rho_{\rm vir}(z)=\Delta_c(z) \rho_{\rm cr}(z)$ is the virial density, $\rho_{\rm cr}(z)=3H^2(z)/(8\pi G)$ 
is the critical density at $z$, $H(z)$ is the Hubble parameter, and $\Delta_c(z)=18\pi^2+82x-39x^2$ where
$x=\Omega_M(z)-1$.

We then derive the number density of gas by $n(r)=n_{\rm H}(r)+n_{\rm He}(r)$. Here 
$n_{\rm H}(r) = f_{\rm H}\rho(r)/m_{\rm H}$ is the number density of hydrogen, where
$f_{\rm H}=0.739$ is the hydrogen mass fraction and $m_{\rm H}$ is the mass of hydrogen atom.
Similarly,  $n_{\rm He}(r) = (1-f_{\rm H})\rho(r)/m_{\rm He}$ is the number density of helium, where
$m_{\rm He}$ is the mass of helium atom.
Also, the temperature-density relation $T(n)$ and the $\rm H_2$ fraction-density relation 
$f_{\rm H_2}(n)=n_{\rm H_2}/n$ can be derived from existing numerical simulations. Here we use the results on
$T(n)$ and $f_{\rm H_2}(n)$ from Omukai (2001) and Yoshida et al. (2006), which are available for
$n\simeq 10^{-2}$ to $10^{23}\ \rm cm^{-3}$ as shown in the left panel of Fig.~\ref{fig:H2_gas}.
The uncertainties of the gas temperature and $\rm H_2$ fraction are shown in blue regions.
These uncertainties are evaluated based on the 
differences in the far-ultraviolet radiation background from the first stars and quasars 
(Omukai 2001).

As can be seen, the gas temperature does not monotonously increase with the gas density. For instance, 
it drops from $T\sim 2000$ to $200$ K between $n\simeq 1$ and $10^4$ $\rm cm^{-3}$ where molecular
hydrogen density  is rising to $f_{\rm H_2}\sim 10^{-3}$. This indicates that  $\rm H_2$ cooling is
starting to become important in this gas density range. 
At $n\simeq 10^4$ $\rm cm^{-3}$, $\rm H_2$ cooling saturates and 
turns into the cooling at the LTE. For $n= 10^{10}\sim 10^{11}$ $\rm cm^{-3}$, almost all of gas particles become 
molecular hydrogen due to the efficient $\rm H_2$ three-body reaction (Yoshida et al. 2006), and we find
$f_{\rm H_2}\simeq 0.5$ by definition. At $n\simeq 10^{20}$ $\rm cm^{-3}$ with $T\simeq 10^4$ K, 
$\rm H_2$ begins to dissociate and the fraction drops quickly to $f_{\rm H_2}<10^{-5}$ when 
$n\simeq 10^{23}$ $\rm cm^{-3}$ and $T\simeq 10^5$ K.

Next, with the help of Eq.~(\ref{eq:p_gas}), we can evaluate the gas temperature and $\rm H_2$ fraction
as a function of the halo radius, i.e. $T(r)$ and $f_{\rm H_2}(r)$. Once these are established,
we can derive $n_{\rm H_2}(r)$, $n_{\rm H}(r)$ and $\Lambda_{\rm H_2}^{\rm r,v}(r)$ which 
are needed for the H$_2$ luminosity calculation. 
In the right panel of Fig.~\ref{fig:H2_gas}, we show the density profile of the gas 
and molecular hydrogen in blue solid and dashed lines for a dark matter halo with $M=10^6\ \rm M_{\sun}/h$ at $z=15$. 
The blue region shows the uncertainty of the H$_2$ density profile which is derived by the uncertainty
of $f_{\rm H_2}$ in the left panel of Fig.~\ref{fig:H2_gas}.
The gas density profile is a straight line with a slope of -2.2 as
indicated by Eq.~(\ref{eq:p_gas}). On the other hand, the density profile of molecular hydrogen has
a more complex shape which is dependent on the relation between $f_{\rm H_2}$ and gas density $n$. For the
outer layer of the gas cloud ($r>10^{-2}$ pc), gas density is less than $10^{5}\ \rm cm^{-3}$ and
$f_{\rm H_2}\lesssim 4\times 10^{-4}$. Here $\rm H_2$ density is much smaller than the gas density.
For the inner region with $10^{-3.5}<r<10^{-2}$ pc, we find $10^{5}<n<10^{11}\ \rm cm^{-3}$ and 
$f_{\rm H_2}$ begins to rise up quickly with $n_{\rm H_2}$ becoming close to the gas density.
For the inner-most region at $r<10^{-3.5}$ pc, the gas density is greater than $10^{11}\ \rm cm^{-3}$, 
and $f_{\rm H_2}\simeq 0.5$ so that almost all of hydrogen end up forming molecular hydrogen.

The luminosity of the $\rm H_2$ rotational or vibrational lines can then be estimated by
\ba
L_{\rm H_2}^{\rm r,v}(M,z) &=& 4\pi \int_0^{r_{\rm vir}}dr r^2 n_{\rm H_2}^{\rm r,v}(r) \nonumber \\
                         &\times&\left[n_{\rm H}(r)\Lambda_{\rm H_2}^{\rm r,v}(\rm H)+n_{\rm H_2}(r)\Lambda_{\rm H_2}^{\rm r,v}(\rm H_2)\right],
\ea
where $n_{\rm H_2}^{\rm r,v}(r)$ is the number density of the molecular hydrogen that can emit at a given
rotational or vibrational line at $r$. We first evaluate the total $n_{\rm H_2}$ at v=0 and 1 states
by condensing all the rotational levels at a given vibrational state to be a single vibrational level,
$n_i=n_{i-1}\ {\rm exp}[-\Delta E_{i,i-1}/(kT)]$ where $i=1$. Here $g_{\rm 0}=g_{\rm 1}=1$ 
for v=0 and 1, respectively, 
and $\Delta E_{10}/k \simeq 5860$ K (Hollenbach \& McKee 1979).
Then we estimate $n_{\rm H_2}$ for a given rotational energy level $J$ in a vibrational level $i$ by 
${n_J} = n_{J'}\ ({g_J}/g_{J'})\ {\rm exp}[-\Delta E_{J,J'}/(kT)]$.  The fractions of the ortho 
and para states of total $n_{\rm H_2}$ are assumed to be 0.75 and 0.25, respectively, in our calculation.

\begin{figure}[htb]
\includegraphics[scale = 0.4]{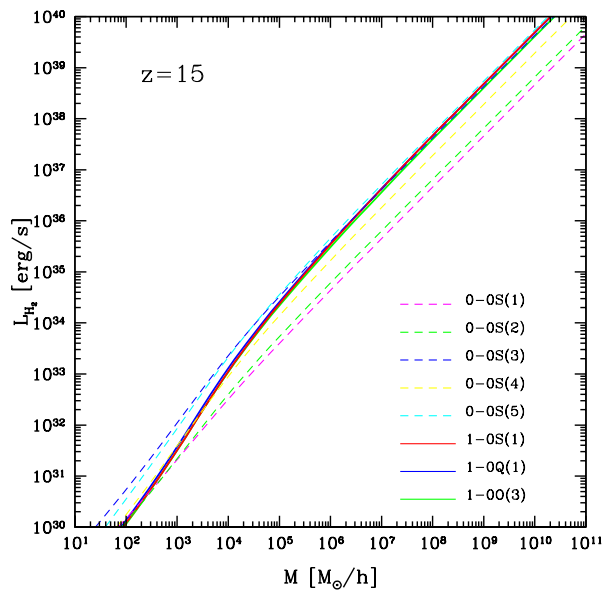}
\caption{\label{fig:H2_lum} $\rm H_2$ line luminosity vs. dark matter halo mass $M$ at $z=15$.
To avoid crowding we select the first eight strongest lines to show here.
In our calculations we find that the 0-0S(3) is the most 
luminous line, while lines such as 0-0S(5), 1-0S(1), 1-0Q(1) and 1-0O(3), are comparable
for high halo masses.}
\end{figure}

In Fig.~\ref{fig:H2_lum}, we show $\rm H_2$ luminosity of several lines as a function
of the halo mass at $z=15$. In these lines, we find that the rotational line 0-0S(3) at a rest-frame wavelength of
$9.66$ $\rm \mu m$ is the most luminous one. Other lines,
such as 0-0S(5), 1-0S(1), 1-0Q(1) and 1-0O(3), are also strong for halos with high mass
(see also Table~\ref{tab:H2_data}).
As can be seen, for low halo masses with $M\lesssim 10^5$ $\rm M_{\sun}/h$, the rotational lines are stronger 
than the vibrational lines. This is caused by the fact that the halos with low masses have lower
mean gas temperature than the massive halos, and the rotational cooling is stronger than the vibrational cooling
in such halos, as indicated by Fig.\ref{fig:H2_cooling}.

\section{$\rm H_2$ intensity and power spectrum}

Given the relation between $\rm H_2$ luminosity and the dark matter halo mass, the mean intensity 
of the $\rm H_2$ lines can be expressed as (Visbal \& Loeb 2010; Gong et al. 2011)
\be
\bar{I}_{\rm H_2}(z) = \int_{M_{\rm min}}^{\infty}dM \frac{dn}{dM}(M,z)\frac{L_{\rm H_2}(M,z)}{4\pi D_{\rm L}^2}y(z)D_{\rm A}^2 \, ,
\ee
where we choose $M_{\rm min}=10\ M_{\sun}/h$, $dn/dM$ is the halo mass function (Sheth \& Tormen 1999),
$y(z)=d\chi/d\nu=\lambda_{\rm H_2}(1+z)^2/H(z)$ when $\chi$ is the comoving distance
and $\lambda_{\rm H_2}$ is the wavelength of $\rm H_2$ lines in the rest frame. 
Our results are not strongly sensitive to the exact value of minimum halo mass. If we increase the minimum halo mass
to the level of 10$^6$ M$_{\sun}/h$, the mean intensity we present here decrease by a factor of $\sim$ 2 for all
H$_2$ lines.

\begin{figure}[htb]
\includegraphics[scale = 0.4]{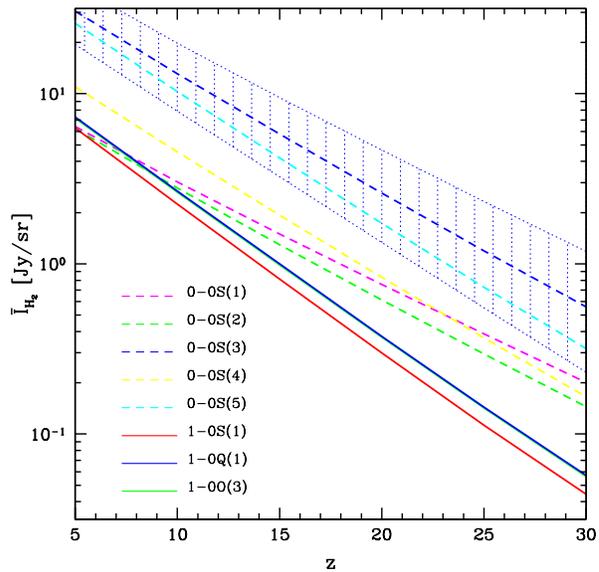}
\caption{\label{fig:Iv_z} The mean intensity of $\rm H_2$ lines as a function of redshift $z$.
The blue region is the uncertainty of the intensity of 0-0S(3) line which is estimated from
uncertainties on the gas temperature and $f_{\rm H_2}$ shown on the left panel of 
Fig.\ref{fig:H2_gas}. We find the 0-0S(3) line is the most luminous line for $10 \le z \le 30$, 
and the slopes of the relations for the vibrational lines are generally steeper than 
the rotational lines. Note that for $ 5 \le z\le 10$ we do not consider the dissociation effect of
the molecular hydrogen by Pop II and Pop III stars.
}
\end{figure}

\begin{figure*}[htbp]
\centerline{
\includegraphics[scale = 0.4]{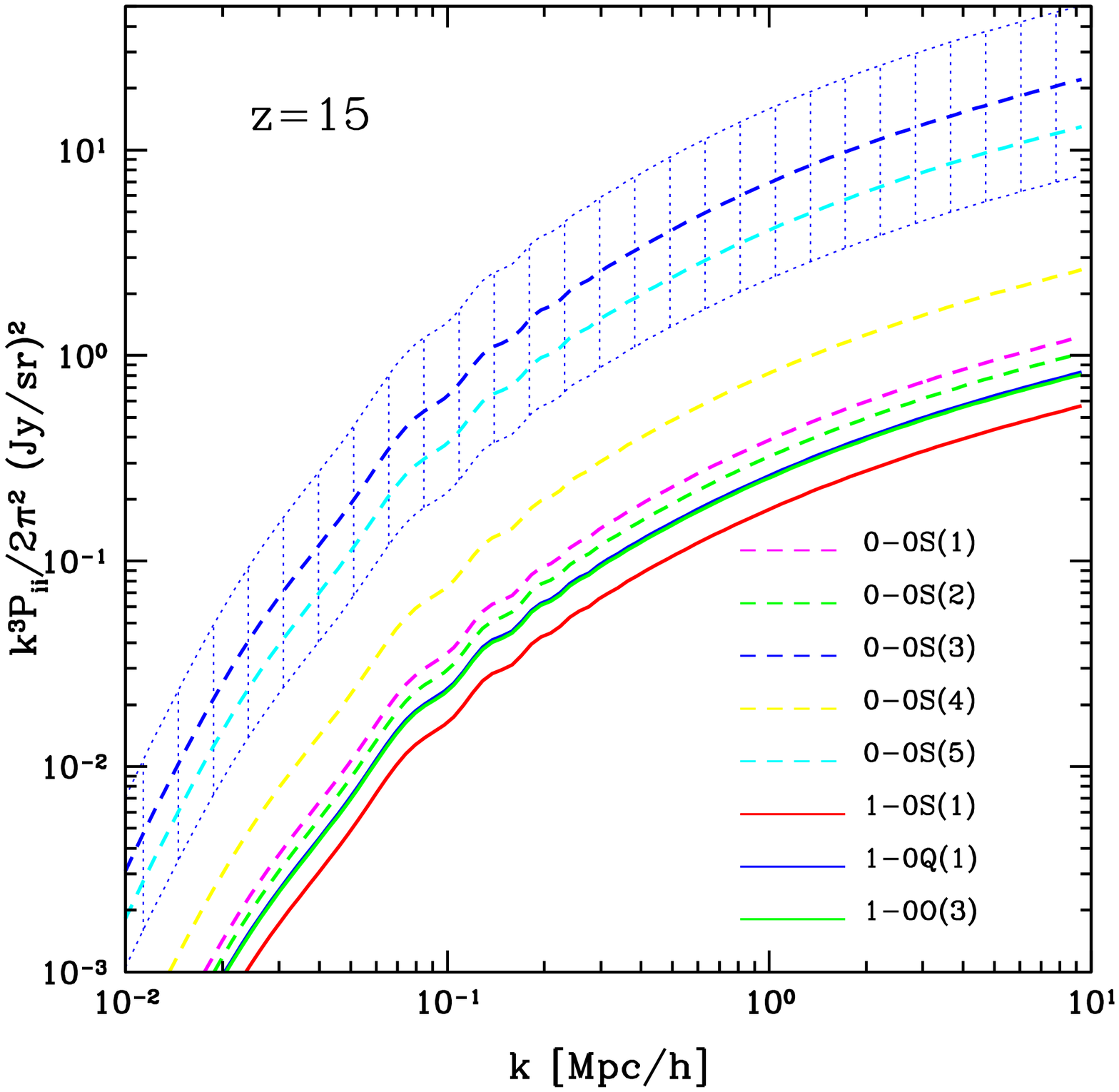}
\includegraphics[scale = 0.4]{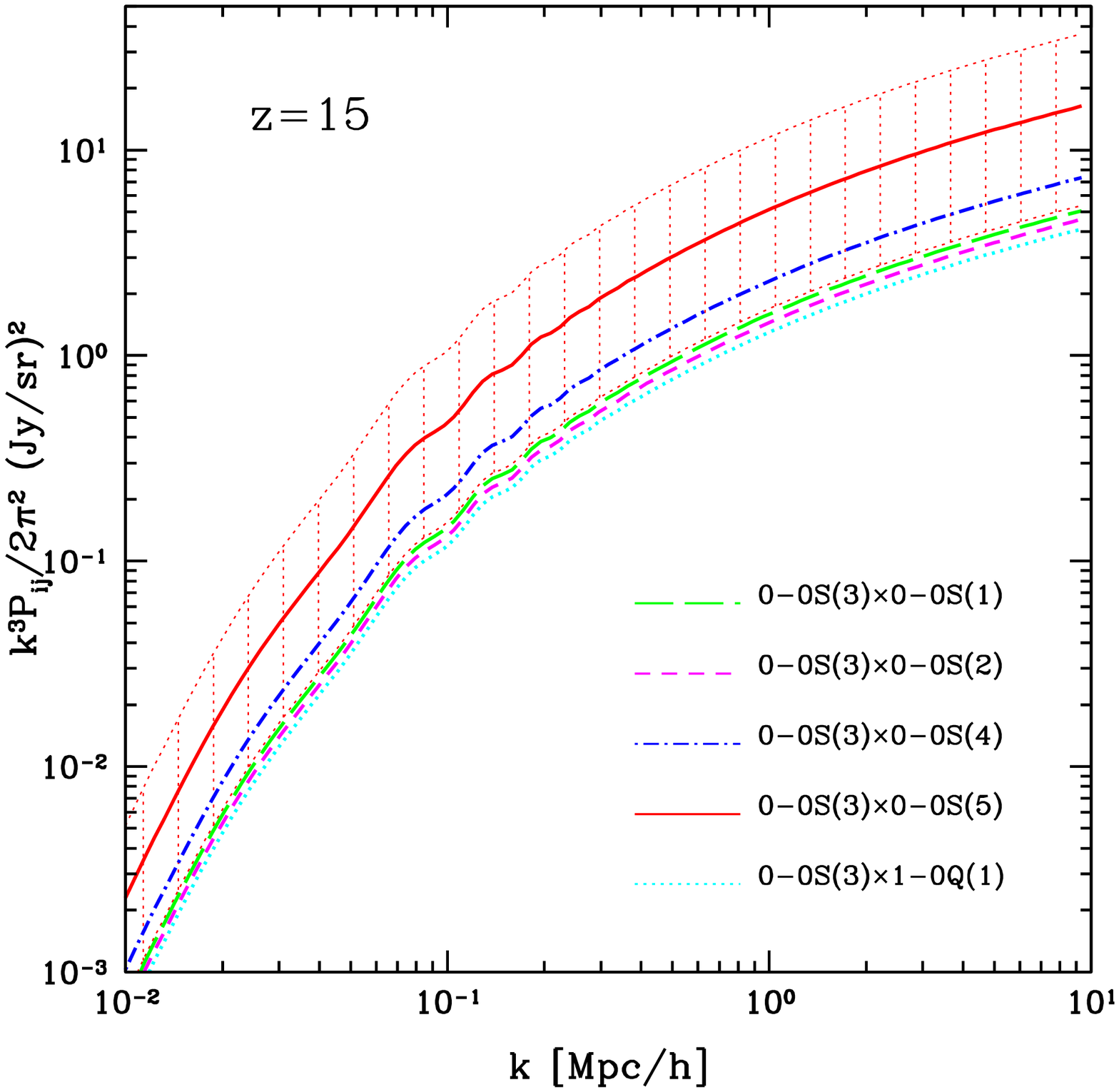} 
}
\caption{\label{fig:PS}  
{\it Left}: The $\rm H_2$ clustering auto power spectrum at $z=15$. Eight brightest
lines are selected to show here. The blue region is the uncertainty on the
clustering power spectrum for the 0-0S(3) line and is derived from the uncertainties
in the gas temperature and $f_{\rm H_2}$.
{\it Right:} The $\rm H_2$ clustering cross power spectrum at $z=15$. 
Here we choose the strongest 0-0S(3) line to cross correlate with  next five strong
lines. The propose cross-correlation effectively eliminates the astrophysical line confusion
from low-redshift sources and other mid-IR lines.
The red region shows the uncertainty for 0-0S(3)$\times$0-0S(5), which is also estimated from the
uncertainties in the gas temperature and $f_{\rm H_2}$.
}
\end{figure*}

In Fig.~\ref{fig:Iv_z}, we show the mean intensity of the eight strongest $\rm H_2$ lines 
as a function of redshift $z$. The uncertainty in the intensity of the 0-0S(3) line is
shown with the shaded blue region, which is derived from uncertainties in the
gas temperature and $f_{\rm H_2}$ in the left panel of Fig.\ref{fig:H2_gas}.
We find that the mean intensity of the 0-0S(3) rotational line is the strongest for 
$10\le z\le30$. This is because the 0-0S(3) line is the most luminous line for low-mass halos, which 
have a higher number density and dominate the halo distribution at $z=15$. 
Also, the slopes of the intensity-redshift relation for the vibrational lines are steeper than that of the rotational 
lines, since they have steeper slopes for cooling coefficients with temperature 
as shown in Fig.~\ref{fig:H2_cooling}. However,
we find the difference in slopes to become 
smaller for the rotational lines as $J$ is increased, indicating that 
the high-$J$ rotational lines have similar slopes with  cooling coefficient when compared to that of 
the vibrational lines.  

We note here that  we do not consider the dissociation effect of the molecular
hydrogen by Pop II and Pop III stars in our calculation. We expect the
 formation of these stars to be important at $z < 10$ and that
there would be significant amount of $\rm H_2$
that should be be dissociated by the UV photons emitting from the first stars.
Thus $\rm H_2$ emission could be suppressed significantly at $z\le 10$. At $z \sim 15$, 
there should still be some dissociation but we ignore it to obtain a safe upper limit estimate on the
expected H$_2$ intensity for experimental planning purposes.

Next we can derive the clustering power spectrum of the $\rm H_2$ lines, writing the intensity as
$I_{\rm H_2}(z)=\bar{I}_{\rm H_2}[1+b_{\rm H_2}\delta({\rm \bf x})]$. 
Here $b_{\rm H_2}$ is the average $\rm H_2$  clustering bias, which can be estimated from
\be
\bar{b}_{\rm H_2}(z) = \frac{\int_{M_{\rm min}}^{\infty}dM \frac{dn}{dM} L_{\rm H_2}b(M,z)}{\int_{M_{\rm min}}^{\infty}dM \frac{dn}{dM} L_{\rm H_2}} \, ,
\ee
where $b(M,z)$ is the bias factor for dark matter halos with mass $M$ at $z$ (Sheth \& Tormen 1999).
The $\rm H_2$ clustering auto power spectrum is then given by
\be
P_{\rm H_2}^{\rm clus}(k,z) = \bar{I}_{\rm H_2}^2 \bar{b}_{\rm H_2}^2 P_{\delta \delta}(k,z),
\ee
where $P_{\delta \delta}(k,z)$ is the matter power spectrum, which is obtained from a halo model 
(Cooray \& Sheth 2002). At a high redshift as $z=15$, the structure of matter distribution is extremely linear
and the 2-halo term dominates the power spectrum.

\begin{table}[!t]
\centering
\caption{The wavelength, $\Delta J= J-J'$, spontaneous emission coefficient $A_J$, mean bias and mean intensity for the $\rm H_2$ rotational and vibrational lines at $z=15$.}
\begin{tabular}{l | c | c | c | c | c }
\hline\hline
   H$_2$ line & $\lambda$ (${\rm \mu m}$) & $\Delta J$ & $A_J (\rm s^{-1})$ & $\bar{b}_{\rm H_2}$ & $\bar{I}_{\rm H_2}$ (Jy/sr)\\
\hline 
   0-0S(0) & 28.2 & +2 & $2.94\times10^{-11}$ & $2.6^{+0.4}_{-0.1}$ & $0.08^{+0.28}_{-0.06}$ \\ 
   0-0S(1) & 17.0 & +2 & $4.76\times10^{-10}$ & $2.8^{+0.3}_{-0.3}$ & $1.52^{+4.87}_{-0.83}$ \\
   0-0S(2) & 12.3 & +2 & $2.76\times10^{-9}$ & $3.0^{+0.2}_{-0.3}$ & $1.32^{+2.20}_{-0.61}$ \\
   0-0S(3) & 9.66 & +2 & $9.84\times10^{-9}$ & $3.1^{+0.2}_{-0.2}$ & $5.90^{+3.60}_{-2.64}$ \\
   0-0S(4) & 8.03 & +2 & $2.64\times10^{-8}$ & $3.2^{+0.2}_{-0.1}$ & $1.97^{+0.94}_{-0.92}$ \\
   0-0S(5) & 6.91 & +2 & $5.88\times10^{-8}$ & $3.3^{+0.2}_{-0.2}$ & $4.26^{+2.15}_{-2.1}$ \\
   0-0S(6) & 6.11 & +2 & $1.14\times10^{-7}$ & $3.4^{+0.2}_{-0.2}$ & $0.78^{+0.52}_{-0.40}$ \\
   0-0S(7) & 5.51 & +2 & $2.00\times10^{-7}$ & $3.5^{+0.2}_{-0.3}$ & $1.05^{+1.02}_{-0.54}$ \\
   0-0S(8) & 5.05 & +2 & $3.24\times10^{-7}$ & $3.6^{+0.3}_{-0.6}$ & $0.13^{+0.22}_{-0.07}$ \\
   0-0S(9) & 4.69 & +2 & $4.90\times10^{-7}$ & $3.8^{+0.2}_{-0.9}$ & $0.13^{+0.44}_{-0.07}$ \\
   0-0S(10) & 4.41 & +2 & $7.03\times10^{-7}$ & $4.0^{+0.2}_{-1.3}$ & $0.01^{+0.11}_{-0.01}$ \\
   0-0S(11) & 4.18 & +2 & $9.64\times10^{-7}$ & $4.2^{+0.2}_{-1.6}$ & $0.01^{+0.23}_{-0.01}$ \\
   1-0S(0) & 2.22 & +2 & $2.53\times10^{-7}$ & $3.4^{+0.2}_{-0.7}$ & $0.24^{+0.87}_{-0.12}$ \\
   1-0S(1) & 2.12 & +2 & $3.47\times10^{-7}$ & $3.5^{+0.2}_{-0.7}$ & $0.83^{+3.06}_{-0.42}$ \\
   1-0Q(1) & 2.41 & 0 & $4.29\times10^{-7}$ & $3.4^{+0.2}_{-0.5}$ & $1.00^{+1.96}_{-0.50}$ \\
   1-0O(3) & 2.80 & -2 & $4.23\times10^{-7}$ & $3.4^{+0.2}_{-0.5}$ & $0.99^{+1.96}_{-0.50}$ \\
\hline \hline
\end{tabular}
\label{tab:H2_data}
\end{table}

\begin{figure*}[htbp]
\centerline{
\includegraphics[scale = 0.4]{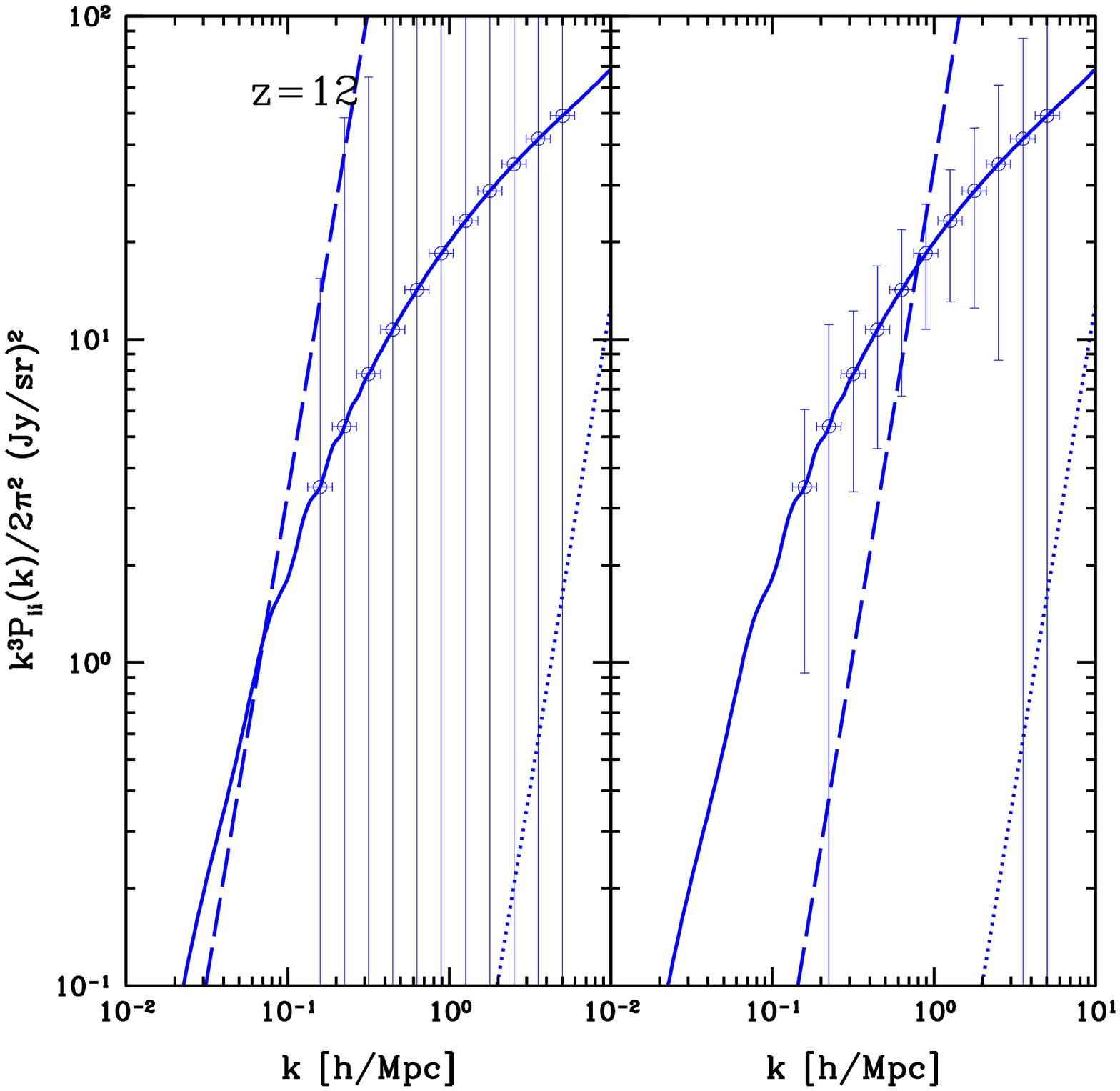}
\includegraphics[scale = 0.4]{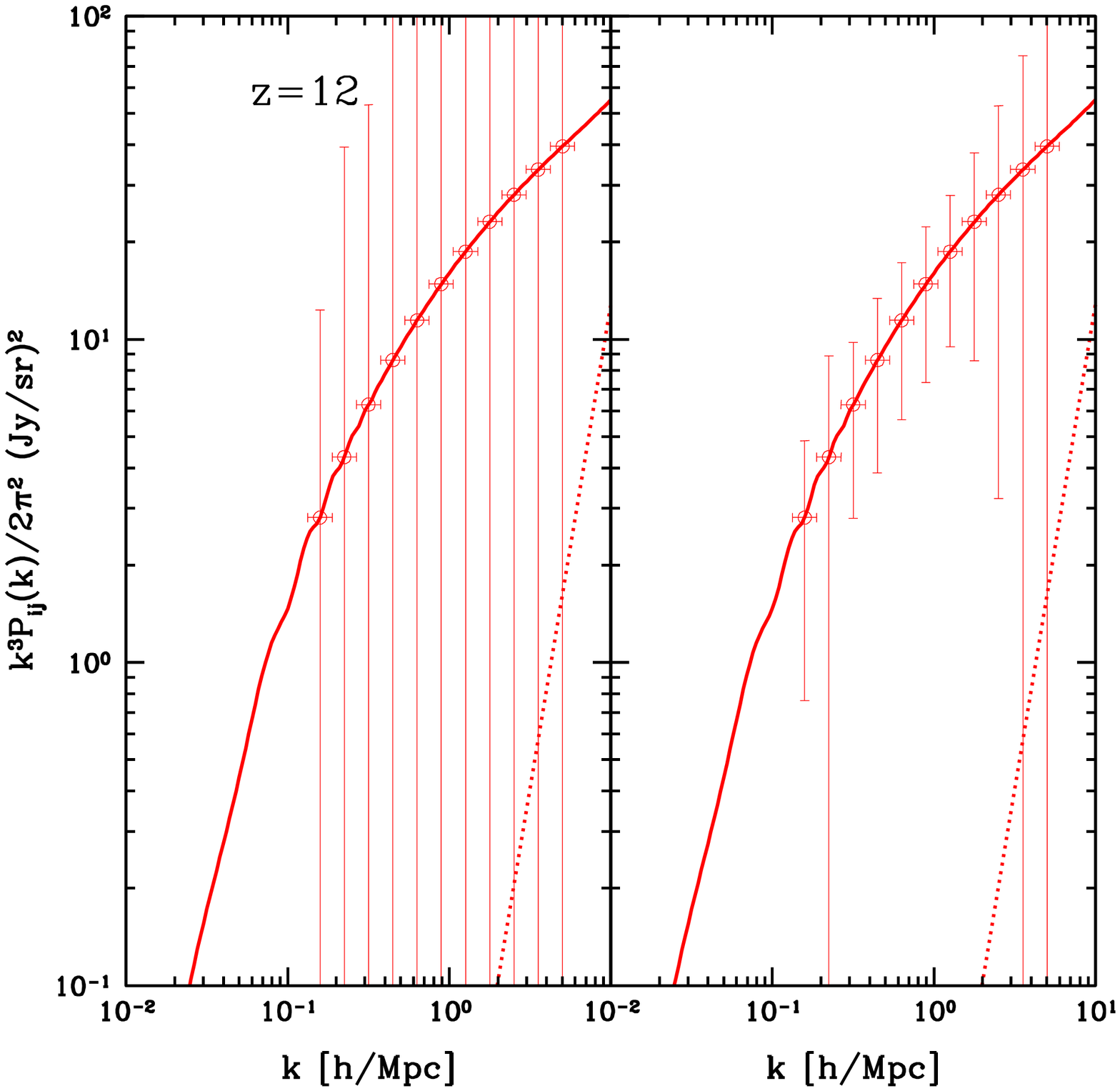} 
}
\caption{\label{fig:PS_err}  
The auto power spectrum of the 0-0S(3) line (left panel) and the cross power spectrum of the 
0-0S(3)$\times$0-0S(5) (right panel) at $z=12$ with the errors estimated for a SPICA/BLISS-like and $10\times$
better SPICA/BLISS-like surveys in each plot. The noise power spectrum and shot-noise power spectrum 
are shown in long-dashed and dotted lines, respectively.
}
\end{figure*}

We can also estimate the shot-noise power spectrum for the $\rm H_2$ lines, which is caused by the 
discretization of the spacial distribution of the primordial clouds, 
\be
P_{\rm H_2}^{\rm shot}(z) = \int_{M_{\rm min}}^{\infty} dM \frac{dn}{dM} \left[ \frac{L_{\rm H_2}}{4\pi D_{\rm L}^2} y(z) D_{\rm A}^2 \right]^2 \, .
\ee

In Table~\ref{tab:H2_data}, we tabulate the rest-frame 
wavelength, $\Delta J=J-J'$, spontaneous emission coefficient $A_J$, 
mean bias and mean intensity for 13 rotational and 4 vibrational lines at $z=15$. The uncertainties of the mean
bias and intensity are evaluated by the uncertainty of the gas temperature and $f_{\rm H_2}$ from the simulations.
We find that the mean intensity of the 0-0S(3) 
rotational line at a rest wavelength of 
$9.66$ $\rm \mu m$ is the strongest among these lines, with a value of around $6$ Jy/sr and a range from
3 to 10 Jy/sr. The other rotational lines such as 0-0S(5), 0-0S(4), 0-0S(1) and 0-0S(2) are 
also bright with total intensities of $\sim$
4.3, 2.0, 1.5 and 1.3 Jy/sr, respectively, at $z=15$. 
The vibrational lines  1-0S(1), 1-0Q(1) and 1-0O(3) have low mean intensities at the level of
0.83, 1.0 and 0.99 Jy/sr, respectively. 
The mean bias factor of these lines lies between 2.6 (for 0-0S(0)) and 4.2 (for 0-0S(11)),
and the bias factors of the rotational lines at higher rotational energy level are higher than that at lower level. 
This is because the lines with high $J$ are stronger at higher mass halos where the temperature is larger.

In the left panel of Fig.~\ref{fig:PS}, the clustering auto power spectrum of eight $\rm H_2$ lines at $z=15$ 
are shown. We find that the shot noise power spectrum $P_{\rm shot}$ is relatively small
compared to the clustering power spectrum $P_{\rm clus}$, and would not affect the $P_{\rm clus}$ at the scales of
interest. This is easy to understand if we notice that the halo mass function is dominated by halos with low masses 
which are more abundant. 

We also calculate the cross correlation between two different $\rm H_2$ lines. Such a cross-correlation will
reduce the astrophysical 
contamination from the other sources, such as low-redshift emission lines from star-forming galaxies, including
63 $\mu$m [OI] and 122 $\mu$m [NII], among others. At $z \sim 15$, the dominant
rotational line 0-0S(3) would be observed at a wavelength of  155 $\mu$m. Such a line would be contaminated
by, for example, $z\sim 0.3$ galaxies emitting [NII]. Thus the auto power spectrum would be higher than
what we have predicted given that the line intensities of [NII] are higher than H$_2$ lines. To avoid this
astrophysical line confusion we propose a cross-correlation between two rotational or rotational and vibrational
lines of the H$_2$ line emission spectrum.

The cross clustering and shot-noise power spectrum for such two $\rm H_2$ lines $i$ and $j$ can be evaluated as
\be
P_{\rm clus}^{ij}=\bar{I}_{\rm H_2}^i \bar{I}_{\rm H_2}^j \bar{b}_{\rm H_2}^i \bar{b}_{\rm H_2}^j P_{\delta \delta}
\ee
and 
\be 
P_{\rm shot}^{ij} = \int_{M_{\rm min}}^{\infty} dM \frac{dn}{dM} \frac{L_{\rm H_2}^i}{4\pi D_{\rm L}^2}  \frac{L_{\rm H_2}^j}{4\pi D_{\rm L}^2}y^i(z) D_{\rm A}^2 y^j(z) D_{\rm A}^2 \, ,
\ee 
respectively.
From these equations, we find that the
the cross power spectrum should have a similar magnitude to the auto power spectrum.
The clustering cross power spectra $P_{ij}^{\rm clus}$ for several $\rm H_2$ lines at $z=15$ are shown in 
the right panel of Fig.~\ref{fig:PS}. We choose the strongest 0-0S(3) line to
cross correlate with the other 5 bright lines, i.e. 0-0S(1), 0-0S(2), 0-0S(4), 0-0S(5) and 1-0Q(1). We find
that the 0-0S(3)$\times$ 0-0S(5) is the largest cross power spectrum since they are brightest two lines. 
At $z \sim 15$, then we would be cross-correlating the wavelength regimes around
110 and 155 $\mu$m. A search for mid-IR lines revealed no astrophysical confusions from low redshifts that
overlap in these two wavelengths at the same redshift. Thus, while low redshift lines will easily dominate the
auto power spectra of H$_2$ lines, the cross power spectrum will be independent of the low-redshift confusions.
In addition to reducing the astrophysical confusions, the cross power spectra also have the advantage that
it can minimize instrumental systematics and noise, depending on the exact design of an experiment.

\section{Detectability}

In this section we investigate the possibility to detect these lines based on current or future instruments. 
We assume a SPICA-like {\footnote{http://sci.esa.int/science-e/www/area/index.cfm?fareaid=105}} survey with 3.5 m 
aperture diameter, 0.1 $\rm deg^2$ survey area, 10 GHz band width, $R$=700 frequency resolution, 100 spectrometers,
and 250 hours total integration time and noise per detector $\sigma_{\rm pix}=10^6\ \rm Jy\sqrt{s}/sr$ at 100 $\mu$m.
Such an instrument corresponds to the latest design of the mid-IR spectrometer, BLISS, from SPICA
(Bradford et al. 2010).

In Fig.~\ref{fig:PS_err}, we show the errors of the auto power spectrum of 0-0S(3) line and cross power spectrum of
0-0S(3)$\times$0-0S(5) at $z$=12 for two cases, a SPICA/BLISS-like and an experiment with 10$\times$ better sensitivity than with the current design of SPICA/BLISS with 
$\sigma_{\rm pix}=10^5\ \rm Jy\sqrt{s}/sr$. The noise power spectrum from the instrument and shot-noise power spectrum 
caused by the discrete distribution of the gas clouds are also shown in long-dashed and dotted lines, respectively. 
We estimate the noise power spectrum and the errors by the same method described in Gong et al. (2012).
We find the signal to noise ratio S/N is 0.2 and 5.2 for the auto power spectrum of 0-0S(3) line in the two cases, and
S/N = 0.1 and 4.5 for the cross power spectrum of 0-0S(3)$\times$0-0S(5). This indicates that the 
current version of SPICA/BLISS does not have the sensitivity to
measure the intensity fluctuation of the H$_2$ lines at a redshift around $z=12$. We find that the noise
requirements suggest an instrument that is roughly 10 times better in detector noise 
than current SPICA/BLISS for a reliable detection.

\begin{figure}[htb]
\includegraphics[scale = 0.4]{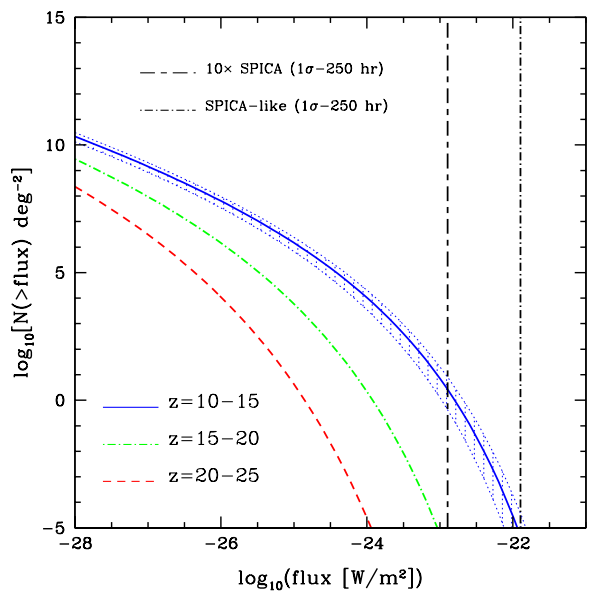}
\caption{\label{fig:N_flux} The number counts of $\rm H_2$ sources for 0-0S(3) line, 
per $\rm deg^2$ with flux greater than a given value for different redshift ranges. 
The uncertainty for $z=10$ to 15 is shown in blue region, which is estimated by the
uncertainties of the gas density and $f_{\rm H_2}$. We also show the flux limit
of a SPICA/BLISS pencil-beam survey 
and another with 10$\times$ better instrumental sensitivity, for a $1\sigma$ detection with 250 hours 
of integration time.}
\end{figure}

In addition to measuring the intensity fluctuations we also explore the detection of the H$_2$ 
point sources at high redshifts.
In Fig.~\ref{fig:N_flux}, we estimate the number of the $\rm H_2$ sources, for 0-0S(3) line per $\rm deg^2$ 
with flux greater than a given value for three redshift ranges $10\le z \le 15$, $15\le z\le20$ 
and $20\le z\le25$. The uncertainty for $10\le z \le 15$ is shown as an example which is derived from the
uncertainties of the simulations. The flux limits of a pencil-beam survey with
 a SPICA/BLISS-like instrument and a 10$\times$ better SPICA/BLISS surveys for $1\sigma$ 
detection with 250 hours of total integration time are also shown in vertical
dash-dotted lines. We find it is hard to detect the 
$\rm H_2$ sources even for the redshift range $10\le z\le 15$ using the SPICA/BLISS-like experiment. The number counts 
of the $\rm H_2$ sources at $10\le z\le 15$ is around $10^{-5}$ per $\rm deg^2$ for the SPICA/BLISS-like survey. For the
instrument that is 10 times better in sensitivity 
than SPICA, we find that, in this first estimate of H$_2$ counts, that we can 
aim to get about 10 sources per $\rm deg^2$ at $10\le z\le 15$.

\section{Discussion and conclusion}

In this paper, we propose intensity mapping of  $\rm H_2$ rotational and vibrational 
lines to detect the primordial gas distribution at large scales during the pre-reionization epochs at $z > 10$. 
At such high redshifts the molecular hydrogen
takes the role of main coolant that leads to the formation of 
first stars and galaxies and the detection of H$_2$ power spectrum can reveal details about the halo mass
scales which first form stars and galaxies in the universe.

We first estimate the cooling rates for both $\rm H_2$ rotational and vibrational lines with the help of fitting
results from Hollenbach \& McKee (1979) and Hollenbach \& McKee (1989). We find the rotational lines 
are dominant at low gas temperature, while the vibrational lines are stronger at high gas temperature. Also, the slope 
of the cooling coefficient-temperature relation for the vibrational lines is steeper than that for the rotational
lines. We then derive the gas number density, temperature and $\rm H_2$ fraction as functions of the halo
radius and estimate the relation of the $\rm H_2$ luminosity and halo mass. 

Next we calculate
the mean intensity for several $\rm H_2$ lines at different redshifts and find the 0-0S(3) is the brightest line 
for $5\le z\le 30$ ($\simeq 5.9$ Jy/sr at $z$=15). Note that we do not consider the dissociation effects of the $\rm H_2$
by the Pop III and Pop II stars at $5\le z\le 10$ that could suppress the H$_2$ emission significantly in this
redshift range. Finally, we evaluate the clustering and shot-noise of auto and 
cross power spectrum for the $\rm H_2$ lines at $z$=15. We find the 0-0S(3)$\times$0-0S(5) is the strongest 
cross power spectrum at $z=15$. We propose such a cross-power spectrum for an experimental measurement
as it has the advantage that it can minimize astrophysical line confusion from low-redshuft galaxies.

In order to consider potential detection of these mid-IR molecular lines
we evaluate the errors of the H$_2$ auto and cross power spectrum at $z=12$ for a SPICA/BLISS-like 
and a design that is 10$\times$ better than the current instrumental parameters. 
We find the S/N for the $z=12$ cross power spectrum detection
is around 0.1 and 5 for these two experiments.
We also estimate the detectability of H$_2$ point sources over different redshift ranges.
We find a SPICA/BLISS-like instrument is not able to detect H$_2$ sources for $z>10$, but
an instrument with 10$\times$ better sensitivity than SPICA/BLISS should be able to
detect about 10 sources per $\rm deg^2$ for $10<z<15$. 
%We ignore that we have presented a first estimate of H$_2$ counts and intensity power spectra. 
We encourage further work on this topic to fully account for
dissociation as stars and galaxies form and additional formation mechanisms, such as shock heating,
that results in H$_2$ line emission from low-redshift galaxies.

\begin{acknowledgments}
This work was supported by NSF CAREER AST-0645427.
MGS acknowledges support from FCT-Portugal under grant PTDC/FIS/100170/2008.
We thank Matt Bradford for helpful discussions and questions that motivated this paper.
\end{acknowledgments}

\appendix

Here we list the fitting formulae of the collisional de-excitation coefficients $C^{\rm H, H_2}_J$ for 
both H$_2$-H and H$_2$-H$_2$ collision from Hollenbach \& McKee (1979) and Hollenbach \& McKee (1989). 
For the rotational cooling in v=0, we have
\be
C_J^{\rm H}(T) = \left( \frac{10^{-11}T_3^{0.5}}{1+60T_3^{-4}}+10^{-12}T_3\right)\left\{0.33+0.9\ {\rm exp}\left[- \left( \frac{J-3.5}{0.9}\right)^2 \right] \right\} {\rm cm^3s^{-1}},
\ee
\be
C_J^{\rm H_2}(T) = (3.3\times 10^{-12}+6.6\times 10^{-12}T_3)\left\{ 0.276J^2\ {\rm exp}\left[ -\left( \frac{J}{3.18}\right)^{1.7}\right] \right\} {\rm cm^3s^{-1}},
\ee
where $T_3=T/1000$ K, and $T$ is the gas temperature. For the vibrational cooling between v=1 and 0, we have
\be
C_{\rm 10}^{\rm H}(T) = 1.0\times 10^{-12}\ T^{0.5}\ {\rm exp}[-(1000/T)]\ {\rm cm^3s^{-1}},
\ee
\be
C_{\rm 10}^{\rm H_2}(T) = 1.4\times 10^{-12}\ T^{0.5} {\rm exp}{-[18100/(T+1200)]}\ {\rm cm^3s^{-1}}.
\ee

The wavenumbers $k=1/\lambda$ of H$_2$ energy levels for $J$=0,...,13 at v=0 and 1 from Dabrowski (1984) are listed in Table \ref{tab:H2_E}. We can get the energy for each level by $E_J=h_{\rm P}ck$, where $h_{\rm P}$ is the Planck constant and $c$ is the speed of light. The wavelengths of the H$_2$ line hence could be derived by $\lambda_{\rm H_2}=h_{\rm P}c/(E_J-E_{J'})$.

\begin{table}[!t]
\centering
\caption{The wavenumbers in $\rm cm^{-1}$ of H$_2$ energy levels for $J$=0,...,13 at v=0 and 1.}
\begin{tabular}{c | c  c}
\hline\hline
   $J$ & v=0 & v=1 \\
\hline 
    0 & 0.00 & 4161.14 \\
    1 & 118.50 & 4273.75 \\
    2 & 354.35 & 4497.82 \\
    3 & 705.54 & 4831.41 \\
    4 & 1168.78 & 5271.36 \\
    5 & 1740.21 & 5813.95 \\
    6 & 2414.76 & 6454.28 \\
    7 & 3187.57 & 7187.44 \\
    8 & 4051.73 & 8007.77 \\
    9 & 5001.97 & 8908.28 \\
    10 & 6030.81 & 9883.79 \\
    11 & 7132.03 & 10927.12 \\
    12 & 8298.61 & 12031.44 \\
    13 & 9523.82 & 13191.06 \\
\hline \hline
\end{tabular}
\label{tab:H2_E}
\end{table}

\end{document}